\documentclass[aps,prc,twocolumn,floatfix,superscriptaddress]{revtex4}
\usepackage{epsfig}
\usepackage{amsmath}
\usepackage{color}

\usepackage{graphicx}

\begin{document}

\title{Triangular flow of thermal photons from an event-by-event hydrodynamic model for 2.76A TeV Pb+Pb collisions at LHC}

\author{Rupa Chatterjee}
\email{rupa@vecc.gov.in}
\author{Dinesh K. Srivastava}
\email{dinesh@vecc.gov.in}
\affiliation{Variable Energy Cyclotron Centre, 1/AF, Bidhan Nagar, Kolkata-700064, India}
\author{Thorsten Renk}
\email{thorsten.i.renk@phys.jyu.fi}
\affiliation{Department of Physics, P.O.Box 35, FI-40014 University of Jyv\"askyl\"a, Finland}
\affiliation{Helsinki Institute of Physics, P.O.Box 64, FI-00014 University of Helsinki, Finland}

\begin{abstract}
We calculate the  triangular flow parameter $v_3$ of thermal photons from an event-by-event ideal hydrodynamic model for 0--40\% central collisions of Pb nuclei at $\sqrt{s_{NN}}$=2.76 TeV at LHC. $v_3$ determined with respect to the participant plane (PP) is found to be non-zero,  positive and its $p_T$ dependence is qualitatively similar to the elliptic flow parameter $v_2$(PP) of thermal photons in the range $1 \le p_T \le 6$ GeV/$c$. In the range $p_T \, \le $ 3 GeV/$c$,  $v_3$(PP) is found to be about 50--75\% of  $v_2$(PP) and for $p_T \, >$ 3 GeV/$c$ the two anisotropy parameters become comparable. The local fluctuations in the initial density distribution as well as the initial global geometry of the produced matter in the event-by-event hydrodynamic framework are responsible for this substantial value of $v_3({\rm PP})$.  However, as expected, the triangular flow parameter calculated with respect to the reaction plane $v_3$(RP) is found to be close to zero. We show that $v_3$(PP) strongly depends on the value of the fluctuation size scale $\sigma$ especially in the higher $p_T \, (\ge 3 {\rm GeV}/c)$ region where a larger value of $\sigma$ results in a smaller $v_3({\rm PP})$. 
 In addition, the $v_3{\rm (PP)}$ is found to increase with the assumed formation time of the thermalized system.

\end{abstract}

\maketitle

\section{Introduction} 

The observation of collective flow and its description in terms of fluid dynamics is a cornerstone of contemporary understanding of the dynamics of ultrarelativistic heavy ion collisions in terms of the creation of thermalized QCD matter.
In many recent studies it has been shown that fluid dynamics utilizing event-by-event (E-by-E) fluctuating initial conditions (IC)~\cite{hannu,pt,scott,hannah, sorenson,nex} can be  successfully used to explain the large elliptic flow results for the most central  collisions and also  the significant triangular flow of charged particles at RHIC and LHC energies ~\cite{alver, flow_phenix,flow_lhc,flow_atlas}  which were underestimated previously by hydrodynamic with smooth initial density distribution.  E-by-E hydrodynamics with fluctuating IC  also explains the hardening of charged particle spectra at larger $p_T$~\cite{hannu,hama}, various structures observed in two particle correlations~\cite{andrade}, and it also has been used to constrain the viscosity over entropy ratio $\eta/s$ from simultaneous measurement of elliptic and triangular flow coefficients~\cite{eta}.

The thermal emission of photons is sensitive to the initial hot and dense stage of the expanding system~\cite{phot} and thus photons are considered as one of the  probes suitable to study IC fluctuations.  It has been shown that fluctuations in the initial density distribution lead to a significant  enhancement in the production of thermal photon compared to a smooth initial density distribution in ideal hydrodynamic calculations~\cite{chre1}. Consequently  a better agreement of the experimental direct photon spectrum is obtained in the region $p_T>$ 2 GeV/$c$~\cite{chre1} using thermal contribution from the fluctuating IC.  The enhancement due to IC fluctuations is found to be more pronounced for peripheral collisions than for central collisions and is less pronounced at the LHC than at RHIC for similar centrality bins~\cite{chre2}.  We have also shown that the elliptic flow calculated using the E-by-E hydrodynamics is substantially larger compared to the results from a smooth initial state averaged profile  in the region  $p_T> 2$ GeV/$c$~\cite{chre3}. 

The success of fluid dynamics implies that the shape of initial spatial geometry or more precisely  the initial spatial anisotropy of the overlapping zone between the two colliding nuclei leads to azimuthally anisotropic pressure gradients. As a result, an anisotropic momentum distribution of the final state particles is observed. The momentum anisotropies of the emitted particles are usually quantified by expanding the invariant particle distribution in transverse plane in terms of the Fourier decomposition
\begin{equation}\label{eq: v2}
 \frac{dN}{d^2p_TdY} = \frac{1}{2\pi} \frac{dN}{ p_T dp_T dY}[1+ 2\, \sum_{n=1}^{\infty} v_n (p_T) \, \rm{cos} (n\phi)] \, .
\end{equation}
Here $\phi$ is the azimuthal angle measured with respect to the reaction plane and various $v_n$ are called the anisotropic flow parameters. The elliptic flow parameter $v_2$ is a result of the almond shape of the initial geometry and  is one of the key observables studied at the RHIC experiments~\cite{fl2}. The significantly  large  $v_2$ measured at RHIC  is considered as a sign of collectivity in the produced system. 


The observation of significant triangular flow anisotropy $v_3$ of hadrons  is attributed to the  collision geometry fluctuations  leading to a potential  initial triangularity of the overlapping zone~\cite{alver}. This is different from the case of $v_2$ where the global shape of the initial collision geometry already has an ellispoid shape which dominates over the local fluctuations. As a result, one finds large $v_2$ with respect to the reaction plane, whereas $v_3$ only takes a non-zero value when determined with respect to an E-by-E determined participant plane.

Triangular flow of photons is of particular interest as photons are emitted from every phase  of the expanding system and at high $p_T$ predominantly reflect early time dynamics, hence they can be expected to particularly probe the initial conditions.
A recent study~\cite{v3_uli} has investigated triangular flow and  higher flow harmonics of thermal photons from an E-by-E viscous hydrodynamic model with KLN and Glauber initial conditions and argued that this leads to a very sensitive measurement of viscosity.

There are three potential mechanisms by which the initial state might influence photon $v_3$: a) fluctuating IC lead to an overall triangular deformation of the matter distribution in the transverse plane which is mapped into a triangular flow patterm. This is the mechanism leading to hadron $v_3$, but since the global shape of the matter distribution can only be probed by late time dynamics when photon production above 1 GeV is suppressed by thermal factors, it is not evident that this mechanism also holds for e.m. emission. b) fluctuating IC lead to irregularly-shaped hotspots with copious photon production, and the early time evolution of such hotspots will likely lead to angular anisotropies in the photon emission. However, there is no obvious reason for this anisotropy to correlate with the hadronic $v_3$ event plane, and hence it can a priori be expected to average out in measurements unless a photon triangular event plane can be determined (which is however impossible in any real measurement due to the very limited number of high $p_T$ photons in any given event) c) hotspots in the fluctuating IC leads indirectly, e.g. by development of additional radial flow, to a magnification of the photon $v_3$. Interpreting any result requires to carefully disentangle these mechanisms.

In the present work we study the $p_T$ dependent triangular flow of thermal photons from fluctuating IC in detail  for 0--40\% central collision of Pb nuclei at LHC and the dependence of $v_3$ on the fluctuation size parameter and the initial formation time of the system and interpret out findings in the light of the three mechanisms outlined above.

\section{Triangular flow of thermal photons from E-by-E hydrodynamic framework}
We use the E-by-E ideal hydrodynamic model framework developed in~\cite{hannu}  to calculate the triangular flow anisotropy of thermal photons  at LHC energy. This (2+1) dimensional hydrodynamic  model  with fluctuating IC has been successfully used to reproduce the $p_T$ spectra and elliptic flow of hadrons at RHIC~\cite{hannu}. It has also been used to calculate the spectra and elliptic flow of thermal photons at the RHIC and at the LHC energies~\cite{chre1,chre2,chre3}. 

A Monte Carlo Glauber model is used for the initial state. The standard two-parameter Woods-Saxon  nuclear density profile is used to randomly distribute the nucleons into the two colliding nuclei. Two nucleons from different nuclei are assumed to collide when the relation $d^2 < \frac{\sigma_{NN}}{\pi^2}$ is satisfied where, $d$ is the transverse distance between the colliding nucleons and $\sigma_{NN}$ is the inelastic nucleon nucleon cross-section. We take $\sigma_{NN}=$ 64 mb at LHC. 

An entropy initialized wounded nucleon (sWN) profile is used where the initial entropy density is distributed around the collision participants (wounded nucleons) using a 2-dimensional Gaussian distribution function
\begin{equation}
  s(x,y) = \frac{K}{2 \pi \sigma^2} \sum_{i=1}^{\ N_{\rm WN}} \exp \Big( -\frac{(x-x_i)^2+(y-y_i)^2}{2 \sigma^2} \Big).
 \label{eq:eps}
\end{equation}
K is an overall constant in the Eq. above and the position of the $i$th nucleon in the transverse plane is denoted by ($x_i, y_i$). 
$\sigma$ is the most important parameter in the above equation which decides the granularity or the size of the initial density fluctuations. It is a free parameter and we use a default value of  $\sigma=$ 0.4 fm as before~\cite{hannu,chre1,chre2,chre3}. We use an initial time $\tau_0=$ 0.14 fm/$c$~\cite{phe} at LHC from the EKRT minijet saturation model~\cite{ekrt} as default value but later vary the value of $\sigma$ and $\tau_0$ from their default values to understand the effect of initial collision geometry and the hotspots on the triangular flow parameter better. 

The temperature at freeze-out is taken as 160 MeV which reproduces the measured $p_T$ spectra of charges pions at LHC. 170 MeV is considered as the transition temperature  from the plasma phase to hadronic phase and we use a lattice based equation of state~\cite{eos} for our calculation. 

We use complete leading order (LO) plasma rates from~\cite{amy} and hadronic rates from~\cite{trg} to calculate the triangular flow of thermal photons from the fluctuating IC. Next-to-leading order (NLO) plasma rates from~\cite{nlo_thermal} are used to obtain the thermal photon spectra from smooth IC only. The total thermal emission  is obtained by integrating the emission rates ($R=EdN/d^3pd^4x$) over the space-time history of the fireball as
\begin{equation}
E \frac{dN}{d^3p}= \int d^4x \, R \left(E^*(x),T(x)\right).
\end{equation}
Here T(x) is the local temperature and $E^* (x)$ = $p^\mu u_\mu (x)$, where $p^\mu$ is the four-momentum of the photons and $u_\mu$ is the local four-velocity of the flow field.
\begin{figure}
\centerline{\includegraphics*[width=8.0 cm]{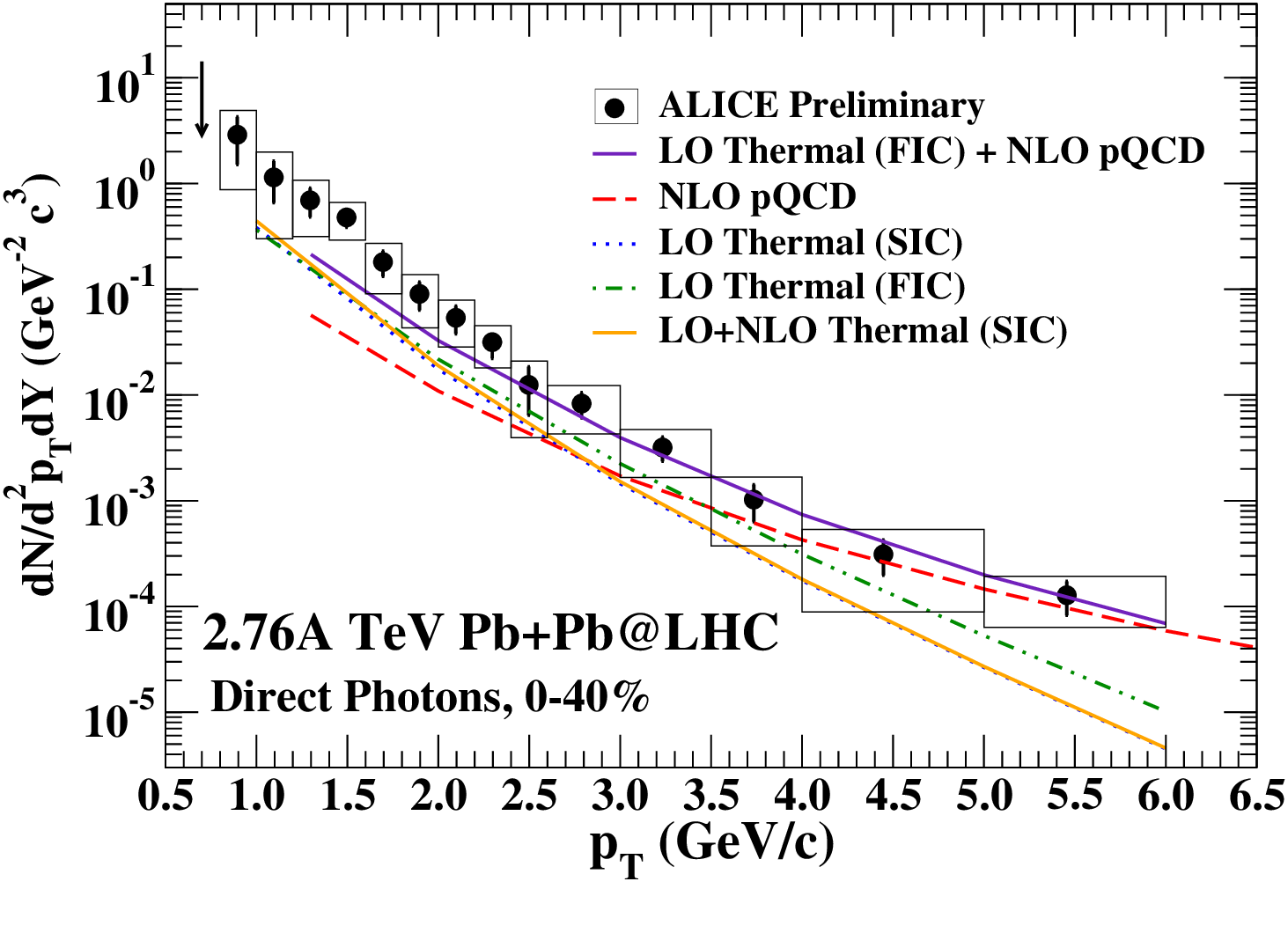}}
\caption{(Color online) Thermal photon $p_T$ spectra considering complete leading order plasma rates~\cite{amy} and next-to-leading order plasma rates~\cite{nlo_thermal} for 2.76A TeV Pb+Pb collisions at LHC and for 0--40\% centrality bin. ALICE direct photons spectrum~\cite{alice} and next-to-leading order pQCD photons are also plotted for comparison.}
\label{spec_lhc}
\end{figure}
The triangular flow parameter $v_3$ is calculated with respect to the participant plane (PP) angle using the relation:

\begin{equation}
v_3^\gamma\{\text{PP}\}=  \langle \cos (3(\phi - \psi_{3}^{\text{PP}})) \rangle_{\text{events}} \, ,
\end{equation}
where the participant plane angle is defined as~\cite{ndhh}

\begin{equation}
  \psi_{3}^{\text{PP}} = \frac{1}{3} \arctan 
  \frac{\int \mathrm{d}x \mathrm{d}y \; r^3 \sin \left( 3\phi \right) \varepsilon\left( x,y,\tau _{0}\right) } 
  { \int \mathrm{d}x \mathrm{d}y \; r^3 \cos \left( 3\phi \right) \varepsilon\left( x,y,\tau _{0}\right)}  + \pi/3 \, .
\end{equation}

Here $\varepsilon$ is the energy density, $r^{2}=x^{2}+y^{2}$, and $\phi$ is
the azimuthal angle. The triangularity or the initial triangular eccentricity of the matter density is calculated using the relation


\begin{equation}
  \epsilon_{3} = -\frac{\int \mathrm{d} x \mathrm{d} y \; r^{3} \cos \left[
  3\left( \phi -\psi_{3}^{\text{PP}}\right) \right] \varepsilon \left(
  x,y,\tau_{0}\right) } {\int \mathrm{d} x \mathrm{d} y \; r^{3} \varepsilon
  \left( x,y,\tau _{0}\right) } \, . 
\end{equation}

\section{Results}
Thermal photon $p_T$ spectra for 0--40\% central collision of Pb nuclei at $\sqrt{s_{\rm NN}}$=2.76 TeV at LHC are shown in Figure~\ref{spec_lhc}. Results from smooth IC (SIC) using complete LO plasma rates (blue dotted line) and NLO plasma rates (solid orange line) are compared. Here the smooth IC is obtained by taking initial state average of 1000 random events which essentially removes all the fluctuations in the initial density distribution.
We see that the addition of NLO contribution to the complete LO rate increases the thermal photon production by 10--15\% in the range $p_T <2$ GeV/$c$, whereas for $p_T>2$ GeV/$c$  LO and NLO spectra look quite similar and fall on top of each other. In addition, we observe that the additional NLO contribution to the thermal photon production is still not sufficient to match the experimental data in the range $p_T<$2 GeV/$c$.

As shown before~\cite{chre2},  thermal photons from fluctuating IC (FIC) along with the contribution from NLO pQCD~\cite{ilkka} explain the direct photon $p_T$ spectrum measured by ALICE~\cite{alice} well in the region $p_T>$2 GeV/$c$. The result from the fluctuating IC is obtained by taking final state average over the $p_T$ spectra from large number of random events. 

The calculation of the elliptic and the triangular  flow anisotropies of thermal photons  from the E-by-E hydrodynamic model is numerically expensive process. The time taken by the  NLO plasma rates to calculate the $p_T$ spectra is significantly larger than the time taken by the LO plasma rates and also the difference between the two rates is not significant for $p_T>$ 2 GeV/$c$.  For this reason we consider it a good approximation to use the LO rates to calculate the flow anisotropies of thermal photons.

\begin{figure}
\centerline{\includegraphics*[width=8.0 cm]{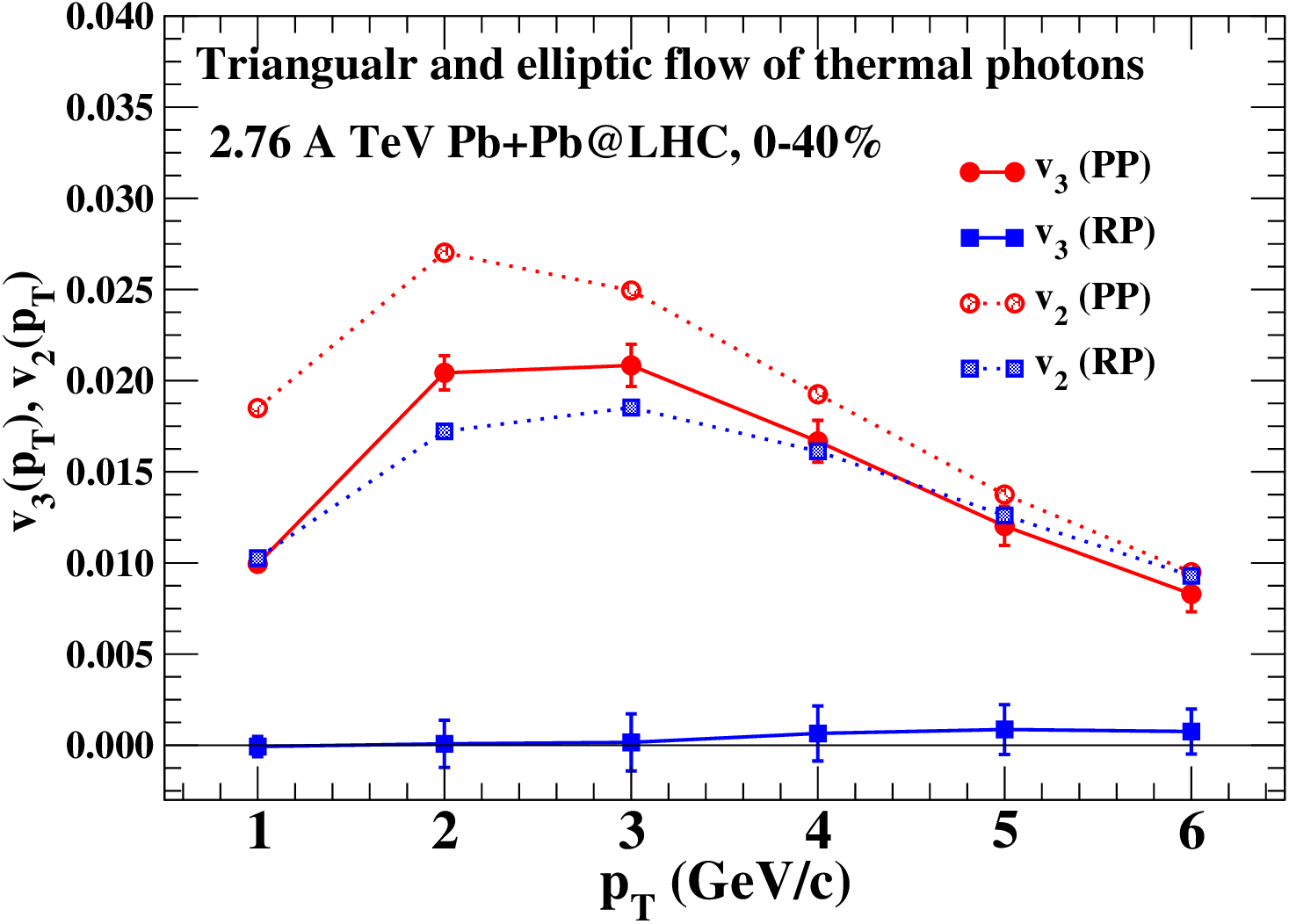}}
\caption{(Color online) Triangular and elliptic flow of thermal photons with respect to RP and PP angles for 0--40\% central collisions of Pb nuclei at LHC and for $\sigma=$0.4 fm.}
\label{v3_lhc}
\end{figure}

Figure~\ref{v3_lhc} shows triangular flow parameter $v_3$ of thermal photons as a function of $p_T$ for 0--40\% central collision of Pb nuclei at $\sqrt{s_{NN}}$=2.76 TeV at LHC. The $v_3$ calculated with respect to the participant plane angle (red solid line closed circle) as well as to the reaction plane (RP) (blue solid line closed squares) are shown for $\sigma$=0.4 fm.  The elliptic flow parameters $v_2({\rm PP})$ and $v_2({\rm RP})$  calculated for the same centrality bin~\cite{chre3} are shown as well for comparison. The $v_3$ results are obtained by averaging over the triangular flow results  from 400 random events and we also show the standard errors on both the $v_3({\rm PP})$ and $v_3({\rm RP})$.

We see that $v_3({\rm PP})$ for thermal photons is positive and significant compared to the the elliptic flow results calculated for the same centrality bin in the region $1 \le p_T \le 6$ GeV/$c$. At $p_T=$ 1 GeV/$c$, the difference between $v_3({\rm PP})$ and $v_2({\rm PP})$ is maximum and $v_3({\rm PP})$ is almost half of the value of $v_2({\rm PP})$ at that $p_T$. The difference reduces more and more towards higher values of $p_T$. $v_3({\rm PP})$ is about 80\% of the value of $v_2({\rm PP})$ at $p_T=$3 GeV/$c$ and at $p_T=5$ GeV/$c$, the difference  between the two results is about 10--15\%.

Triangular flow  calculated with respect to the reaction plane from individual events are found to be both positive and negative. As expected, the averaged $v_3({\rm RP})$ is  zero within standard errors as shown in the Figure~\ref{v3_lhc}. 

In order to understand the individual effects of global geometry and local fluctuations on the triangular flow results better we study  $v_3$  of individual events by keeping the the number of wounded nucleons ($\rm N_{\rm WN}$) fixed.  We take $\rm N_{\rm WN}=$ 200 and generate random events having different triangular flow eccentricity. The number of binary  collisions $N_{\rm coll}$ also varies in an wide range in those events. 
 
\begin{figure}
\centerline{\includegraphics*[width=8.0 cm]{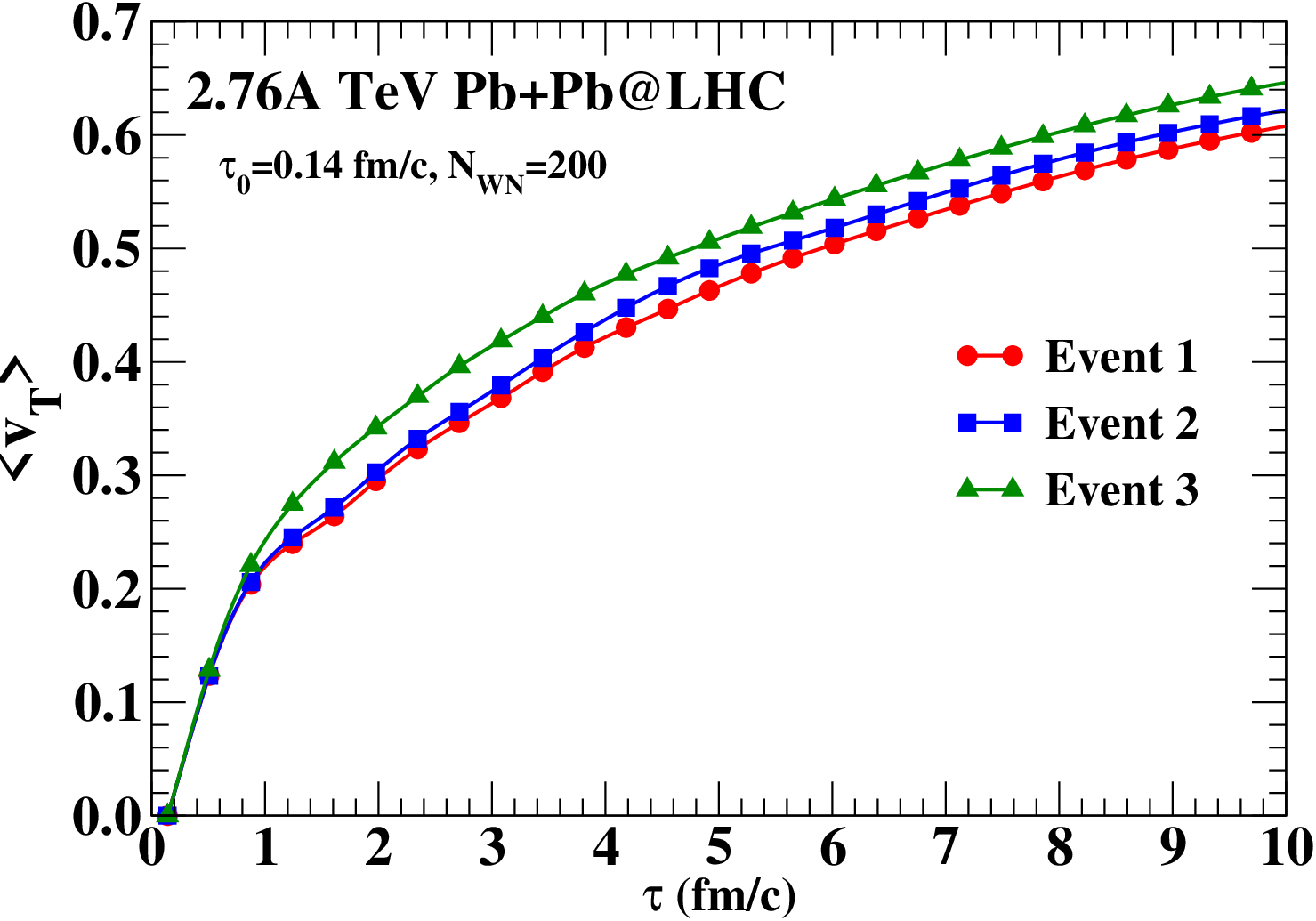}}
\centerline{\includegraphics*[width=8.0 cm]{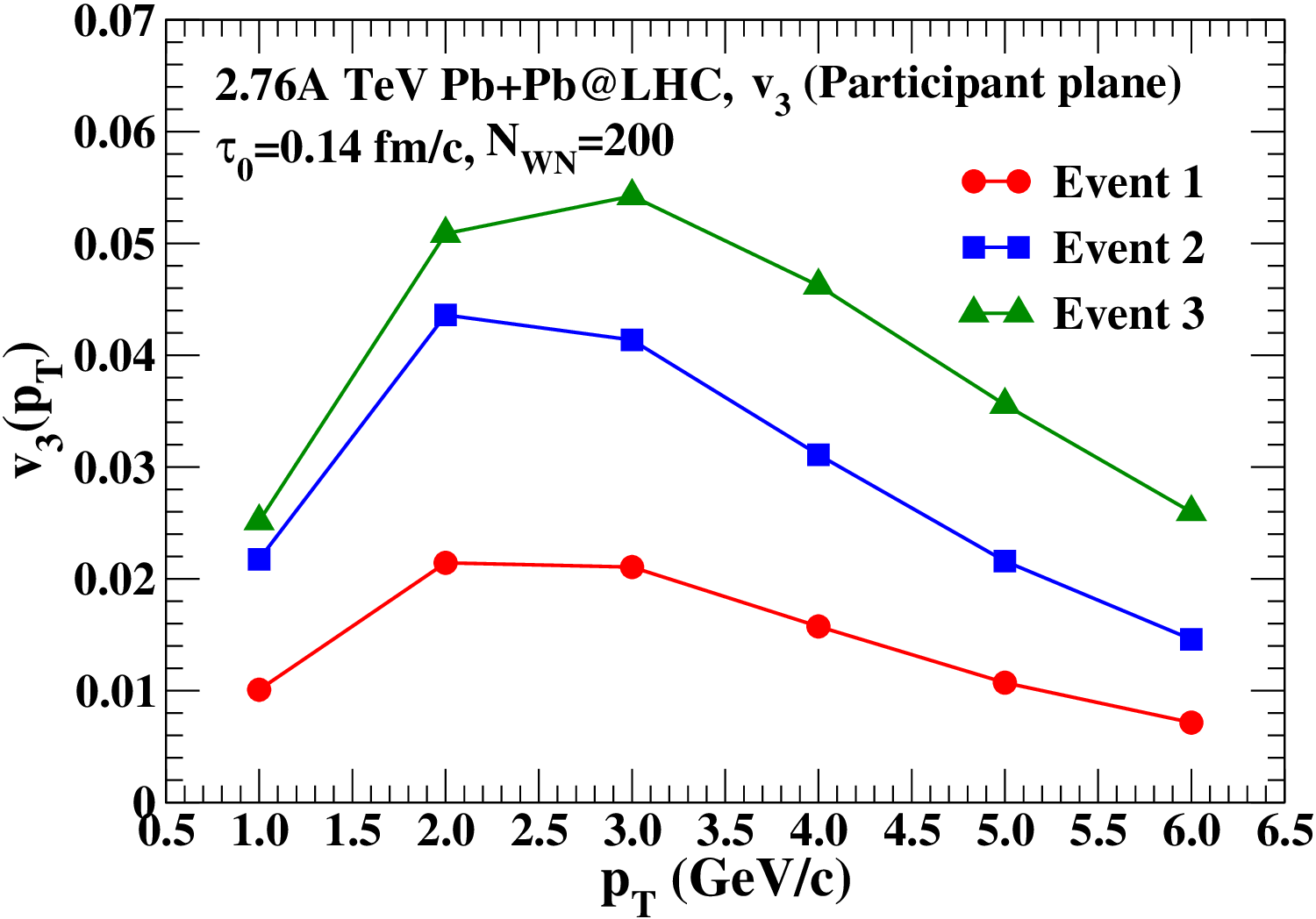}}
\caption{(Color online)\label{F-3} [Upper panel] Average transverse flow velocities for different events with fixed number of wounded nucleons and [Lower panel] $v_3({\rm PP})$ for the same events. }
\label{events}
\end{figure}

The variation of average transverse flow velocity with proper time for three different events with fixed $\rm N_{\rm WN}$ is shown in upper panel of Figure~\ref{events}. The $N_{\rm coll}$ values for Event 1, Event 2, and Event 3 are 537, 611 and 724 respectively and the initial triangular eccentricities of the events are 0.122, 0.325 and 0.177 respectively.  We see that the average transverse flow velocity with time is largest for Event 3, the event with maximum $N_{\rm coll}$ and $\langle v_T \rangle$ is smallest for the Event 1. 

\begin{figure}
\centerline{\includegraphics*[width=8.0 cm]{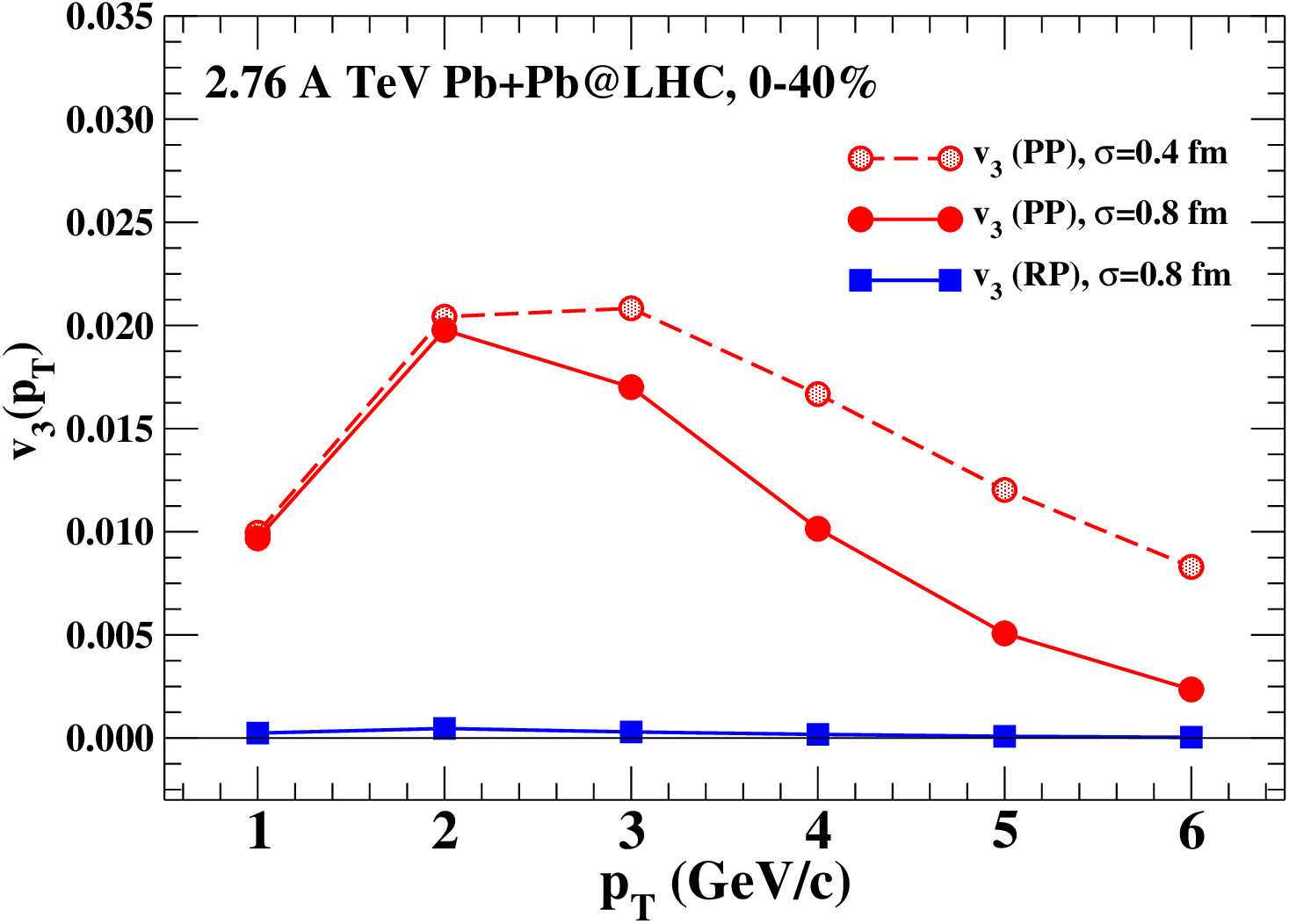}}
\caption{(Color online) Triangular flow of thermal photons for 0--40\% central collisions of Pb nuclei at LHC and for size parameter $\sigma=$ 0.4 and 0.8 fm. }
\label{fig_sigma}
\end{figure}
The $v_3$ calculated with respect to the PP angle for the three events is shown in lower panel of Figure~\ref{events}. $v_3$ shows ordering similar to the average transverse flow velocity where, $v_3(\rm{PP}$) is largest for Event 3 and is smallest for Event 1 although, the triangular eccentricity is maximum for Event 2. This argues that indirect effects of fluctuations, such as the buildup of a strong flow field, contribute significantly to the observed result beyond leading to an overall triangular eccentricity. For hadrons a clear mapping between the initial eccentricity and $v_n$ has been observed~\cite{ndhh}, however this is not true for photons.

%


The fluctuation size scale $\sigma$ is an important parameter as both the geometry and the overal  strength photon emission is strongly sensitive to its value.
The initial density distribution becomes  smoother for larger values of size parameter and in an earlier study~\cite{chre3} we have shown that the elliptic flow parameter for $\sigma = $ 1.0 fm is quite similar to the elliptic flow parameter calculated using a smooth initial state averaged IC.   Thus, it is crucial to study
the dependence of photon $v_3$ on the value of $\sigma$. Figure~\ref{fig_sigma} shows $v_3({\rm PP})$ and $v_3({\rm RP})$ as a function of $p_T$   for 0--40\% central collisions of Pb nuclei at $\sqrt{s_{NN}}$=2.76 TeV at LHC  and for two different $\sigma$ values. As shown in the figure, the value of $v_3({\rm PP})$ for $\sigma=$ 0.4 fm (red dashed line, open circles) is close to the $v_3({\rm PP})$ results for $\sigma=$ 0.8 fm (red solid line, closed circles)  in the lower $p_T \, (\le 2 \, {\rm GeV}/$c$)$ region. However, for $p_T >$ 2 GeV/$c$, $v_3({\rm PP})$ is smaller for larger value of $\sigma$ and with increasing $p_T$, the  $v_3({\rm PP})$  for $\sigma=$ 0.8 fm falls sharply compared to the flow result for $\sigma=$ 0.4 fm.
The presence of the local fluctuations or the hotspots in the IC results in stronger radial flow velocity which allows to probe global geometry more efficiently and we see larger $v_3$ for the smaller value of  $\sigma$.

One can expect even larger $v_3({\rm PP})$ (than the results shown in Figure~\ref{v3_lhc}) when $\sigma$ is smaller than 0.4 fm. However, the flow anisotropy calculation  become numerically expensive with smaller values of $\sigma$ and thus we can not show results for $\sigma <$ 0.4 fm.

As expected, the triangular flow parameter calculated with respect to the RP does not depend on the value of  $\sigma$ and $v_3({\rm RP})$ close to zero for  $\sigma=$ 0.8 fm (shown by black solid line, closed triangles).

Next we study the dependence of the triangular flow parameter on the initial formation time $\tau_0$ of the plasma. All the $v_3$ results shown till now are calculated for a very small $\tau_0=$0.14 fm which is taken from EKRT model~\cite{ekrt} for most central collision of Pb nuclei at LHC. Now the formation time of the plasma can be larger for peripheral collisions than for central collisions~\cite{chre2}. As we do not have the formation time at LHC for  0--40\% centrality bin from EKRT model, we choose a sufficiently large $\tau_0=$ 0.6 fm/$c$ to start the hydrodynamic calculation  in order to see the dependence of $v_3$ on the initial formation time of the plasma.

\begin{figure}
\centerline{\includegraphics*[width=8.0 cm]{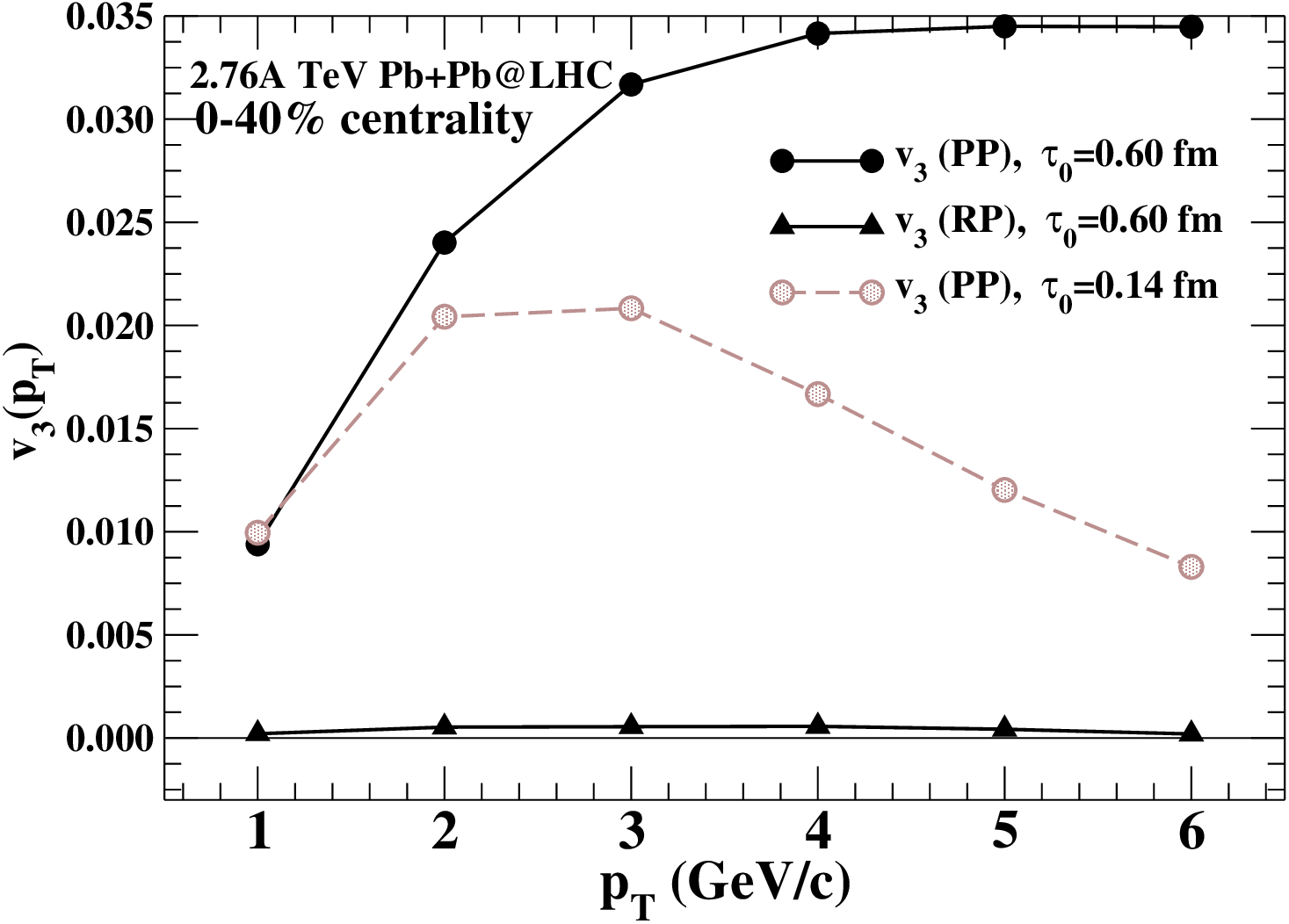}}
\caption{(Color online) Triangular flow of thermal photons for 0--40\% central collisions of Pb nuclei at LHC and for size parameter $\sigma=$ 0.4 and at $\tau_0$ values 0.14 and 0.6 fm/c.}
\label{fig_0.6}
\end{figure}

Figure~\ref{fig_0.6} shows $v_3({\rm PP})$ as a function of $p_T$ for $\tau_0$ values 0.14 and 0.6 fm/$c$ and for $\sigma=$  0.4 fm. It should be noted that the triangular flow parameter for two different $\tau_0$ values is calculated  by keeping the total entropy of the system fixed. 

The $v_3({\rm PP})$ for larger $\tau_0$ rises  rapidly compared to the flow parameter calculated using smaller $\tau_0$ in the region $p_T \le$ 3 GeV/$c$. For $p_T>$ 3 GeV/$c$, $v_3$ for $\tau_0=$ 0.6 fm/$c$  does not change significantly with $p_T$ and becomes constant. This is contrary to the $v_3$ for $\tau_0=$ 0.14 fm/$c$ which drops with increasing $p_T$. The high $p_T$ photons are emitted early when the presence of local fluctuations in the IC is strong. However, these photons do not contribute significantly to the $v_3$ result and the small values of transverse flow velocity in the initial stage brings the $v_3$ down for $p_T >$ 3 GeV/$c$  when a smaller $\tau_0$ is considered.  For $\tau_0=$ 0.6 fm/$c$, a large fraction of these high $p_T$ photons are not included in the calculation and as a result we get much larger $v_3$~\cite{cs}. 

Fig.~\ref{fig_0.6} explicitly shows that the eccentricity of initial hotspots does not contribute to photon $v_3$ significantly as expected, only photons emitted somewhat later carry significant $v_3$. We can combine this with with Fig.~\ref{F-3} to conclude that the most important effect of initial functuations on thermal photon $v_3$ is indirect, i.e. the modification of the radial flow pattern which can transform even a small initial eccentricity into a large $v_3$.

\section{Summary and conclusions}
We calculate the triangular flow anisotropy of thermal photons from an E-by-E ideal hydrodynamic model. 
The complete leading order plasma rates and state of the art hadronic rates are used to calculate $v_3$ at LHC using suitable initial and final conditions.  We show that the inclusion of NLO plasma rate to the complete LO rates does not change the spectra results significantly for $p_T \ge $ 2 GeV/$c$.

The flow parameter $v_3$ as a function of $p_T$ is calculated with respect to the  reaction plane and participant plane angles for 2.76A TeV Pb+Pb collisions at LHC and for 0--40\% central collisions. The value of  $v_3(\rm PP)$ at $p_T=$ 1 GeV/$c$  is found to be about 50\% smaller than the  $v_2(\rm PP)$  calculated using same initial and final conditions. However, the two results become comparable for $p_T >$ 3 GeV/$c$. The triangular flow anisotropy calculated with respect to the RP is close to zero.  

The $v_3(\rm PP)$  calculated using $\sigma=$ 0.8 fm is found to be similar to  $v_3(\rm PP)$ for $\sigma$=0.4 fm in the region $p_T \le $2 GeV/$c$. For larger $p_T$ however, the $v_3$ for larger $\sigma$ is relatively smaller in magnitude.
 We see the variation of average transverse flow velocity  with proper time and $v_3(\rm PP)$ as a function of $p_T$ as well for different events with same number of wounded nucleons. This is done in order to understand the role of $\langle v_T \rangle$ and $\epsilon_3$ in determining the triangular flow anisotropy parameter better. $v_3$ as a function of $p_T$ for two different values of initial formation time is also compared. A larger $\tau_0$ results in a much larger $v_3$ compared to the result calculated using a smaller $\tau_0$.

We find that photon $v_3$ probes the initial state geometry in an indirect way via the generation of additional transverse flow. The sensitivity to the fluctuation size scale as well as to the equilibration time offers useful constraints for a dynamical modeling of the pre-equilibrium phase.

\begin{acknowledgments} 
RC gratefully acknowledge he financial support by the Dr. K. S. Krishnan Research Associateship from Variable Energy Cyclotron Centre, Department of Atomic Energy, Government of India. TR is supported by the Academy researcher program of the Academy of Finland, Project No. 130472. We thank Hannu Holopainen for providing us with the event-by-event hydrodynamic code and for many useful discussions. We also thank Ilkka Helenius for the NLO pQCD photon results and the ALICE collaboration for the direct photon spectrum for 0--40\% central collisions of Pb nuclei at LHC.  We acknowledge the computer facility of the CSC computer centre, Espoo.    
\end{acknowledgments}

\end{document}